\documentclass[aps,pra,twocolumn]{revtex4-2}

\usepackage[usenames,dvipsnames]{color}
\usepackage{graphicx}
\usepackage{amssymb}
\usepackage{subfigure}
\usepackage{comment}
\usepackage{stackrel}
\usepackage{amsmath}
\usepackage{amsfonts}
\usepackage{bm}
\usepackage[normalem]{ulem}
\usepackage{hyperref}
\usepackage{xcolor}
\definecolor{rosita}{rgb}{0.97, 0.56, 0.65}
\newcommand{\ket}[1]{\ensuremath{\left|{#1}\right\rangle}}
\newcommand{\bra}[1]{\ensuremath{\left\langle{#1}\right |}}
\newcommand{\beq}{\begin{equation}}
\newcommand{\eeq}{\end{equation}}
\newcommand{\be}{\begin{equation}}
\newcommand{\ee}{\end{equation}}
\newcommand{\ben}{\begin{eqnarray}}
\newcommand{\een}{\end{eqnarray}}
\definecolor{colvic}{rgb}{0.01, 0.75, 0.24}

\usepackage{graphicx}
\usepackage{color} 
\usepackage{bm}   

\begin{document}


\title{
Enhancing teleportation via noisy channels: effects of the induced multipartite entanglement\\
}

\author{Victor H. T. Brauer}
\email{hernantorres@estudiantes.fisica.unam.mx}

\author{Andrea Valdés-Hernández}%
 \email{andreavh@fisica.unam.mx }
\affiliation{%
 Instituto de Física, 
 Universidad Nacional Autónoma de México \\
 Apartado Postal 20-364, Ciudad de México, Mexico
}%

%


\begin{abstract}
 Quantum teleportation in the presence of noisy channels acting on a bipartite resource state is considered. 
We consider a family of generalized noisy 
 channels that continuously connect the amplitude damping and the dephasing channels, encompassing a wide family of in-between scenarios, to delve into the relation between the teleportation success and the amount of 3- and 4-partite entanglement (distributed among the qubits of the resource state and those representing local environments) generated during the evolution. 
Our analysis reveals that 
for a fixed entanglement of the resource state, the channels that better protect the teleportation fidelity against the detrimental effects of noise are those that generate higher amounts of (\textsc{ghz}-type) multipartite entanglement
This suggests that the dynamically induced multipartite correlations may serve as an additional resource for teleportation, and throws light into the characterization of processes, and of the type of induced entanglement, according to their capability of assisting the protocol. 
\end{abstract}

\maketitle
\section{Introduction}

Quantum teleportation stands out as one of the most fascinating applications of quantum entanglement.
It allows the transmission of quantum information between two spatially separated agents, Alice and Bob, by means of local operations, classical communication, and a key element: a shared correlated state known as  \textit{resource state}
\footnote{Some authors refer also to the resource state as \emph{quantum channel}. Here we will keep the term `quantum channel' to denote a completely positive trace-preserving map, that in the present context gives the reduced dynamics of an open system.}. 
The original standard teleportation protocol \cite{Toriginal} uses a (pure, maximally entangled, two-qubit) Bell state as the resource state. 
Subsequent investigations extended the scheme noticing that there exist resource states that are useful for teleportation yet are neither maximally entangled \cite{BanaszekPRL2001} nor pure, and their relation with violations of Bell inequalities \cite{popescu1994,HORODECKI199621} and discord-like correlations \cite{BUSSANDRI2022} has been discussed. 

A mixed rather than a pure resource state is more realistic, particularly when taking into consideration the interaction of the (Alice and Bob's) entangled pair with its surroundings, resulting in mixing of the resource state. 
There has been extensive research on such \emph{noisy quantum teleportation} schemes \cite{Ohetal2002,Jiang2019,Harraz2022,Im2021,badzia2000,BandyoPRA2002,Fortes2015,Ahadpour2020,FonsecaPRA2019,Laura2014}, e.g., 
resorting to the Lindblad formalism to explore the fidelity of teleportation in terms of decoherence rates \cite{Ohetal2002}, or to determine the optimal Bell resource state under different local Pauli noises \cite{Jiang2019}. Recent methods for protecting teleportation against some decoherence channels have also been advanced \cite{Harraz2022,Im2021}.  
The effect on the teleportation fidelity of different noisy channels acting on the resource state ---typically representing the interaction of Alices's and Bob's particles with additional subsystems (which in the present case are regarded as local environments)---, has been studied considering the Kraus operators corresponding to dissipative interactions via an \emph{amplitude damping} channel \cite{badzia2000,BandyoPRA2002}, together with other paradigmatic noise or decoherence channels on qubits such as \emph{bit flip}, \emph{phase flip}, \emph{depolarizing} \cite{Fortes2015}, and \emph{phase damping} \cite{Ahadpour2020}. 
Also, the teleportation protocol under noisy channels in higher dimensional systems has been explored \cite{FonsecaPRA2019}.
A general theoretical and experimental \cite{Laura2014} 
conclusion that ensues from these investigations is that there exist appropriate channels (acting on suitable initially pure resource states) for which the detrimental effects of noise on the teleportation fidelity are minimal, compared to other noisy channels.

Further extensions of the original teleportation protocol have also been advanced that consider multi-party resource states 
exhibiting some type of multipartite entanglement.
This has led to 
the development of strategies that exploit multipartite entanglement to teleport multiple qubit states, as e.g. in  \cite{Lee2002,GorPLA2003,yeo2006,man2007,RigolinPRA2005}. 
Multi-directional teleportation, allowing quantum information transmission between several agents, has also been explored  \cite{LiIJTP2016,ChouIJTP2018,WangQIP2022,SinghOQ2023}, including the effect of noisy channels \cite{GHZvsWnoisy,Yang_2017,Han_W_2008, Han_GHZ_2008}.

Despite the advances achieved regarding the teleportation success under noisy channels, whether acting on bipartite or multipartite resource states, the relation between the teleportation fidelity and the multipartite entanglement generated among the resource qubits and the environment, has been much less explored.
Such an analysis would allow us to identify the type of processes ---characterized by the type of entanglement they induce---, that favor a more successful teleportation, 
and to possibly explain the fidelity improvement as an effect assisted by the created multipartite entanglement.
In \cite{IshizakaPRA2001,VanhopQIP2019} some progress has been made, by relating the teleportation fidelity with the 3-partite entanglement resulting from the local interaction of one qubit of the resource state with a two-level environment.  
Here we contribute along these lines, by focusing on the standard teleportation protocol in the presence of noisy channels acting on the bipartite resource state. 
We consider a  \emph{generalized} noisy quantum channel that can be continuously transformed from the amplitude damping to the dephasing channel \cite{RuanPRA2018}, and explore the correlation between the \emph{maximal average fidelity} and the amount of \emph{3}- and \emph{4-partite} entanglement, distributed among the pair of qubits that conform the resource state and the qubits that represent the corresponding local environments. 

We first revisit the standard teleportation protocol and the notion of maximal average fidelity $ F_{\max}$ as a quantifier of the teleportation success (Sec. \ref{STP}), and express the latter in terms of the Kraus operators of an arbitrary quantum channel acting on the (arbitrary yet initially pure) resource state, conformed by two qubits $A$ and $B$ (Sec. \ref{FK}). 
After these preliminary sections, we relate the maximal average fidelity \emph{above the classical threshold value}, $\mathbb F_{\max}$, with the \emph{bipartite} entanglement between $A$ and $B$, considering $A+B$ as an ideal closed system (Sec. \ref{bipartite}). 
Assuming then that $B$ interacts with a local, two-level environment $E_B$ via the generalized channel, we investigate the relation between $\mathbb F_{\max}$ and the 3-partite entanglement distributed among $A,B$ and $E_B$ (Sec. \ref{Stripartite}). 
The analysis is extended to the 4-partite case by considering that both $A$ and $B$ locally interact with their respective environments $E_A$ and $E_B$ under independent generalized channels, and a non-trivial correlation between $\mathbb F_{\max}$ and the 4-partite entanglement is disclosed (Sec. \ref{Sfourpartite}). Finally, some concluding remarks are presented (Sec. \ref{Close}).     


\section{The Standard Teleportation Protocol and maximal average fidelity}\label{STP}

The main idea behind the standard teleportation protocol is that Alice wants to send to Bob an arbitrary \emph{input} state, $\rho_\textrm{in}$,  encoded in a qubit $a$ in her possession. 
For this task, they share a pair of qubits $A$ and $B$ (in Alice and Bob's possession, respectively) in an entangled state $\rho_{AB}$, called \textit{resource state}. 
Alice then performs a Bell measurement \cite{Bellmeasure0,Bellmeasure1,Bellmeasure2} on her pair of qubits $a$ and $A$, and communicates the outcome to Bob via a classical channel. 
Upon receiving this information ---and knowing $\rho_{AB}$  \cite{probtel}---, Bob performs a unitary operation $\sigma^{(i)}\in\{\mathsf I_2, \sigma^{x},\sigma^{y},\sigma^{z}\}$ on his qubit $B$, thus putting it into the \textit{output state} $\rho_{\textrm{out}}$ ($\sigma^{x,y,z}$ stand for the Pauli matrices, and $\mathsf I_n$ denotes the $n\times n$ identity operator).

\begin{figure}[ht]
  \begin{center}
    \includegraphics[scale=0.8]{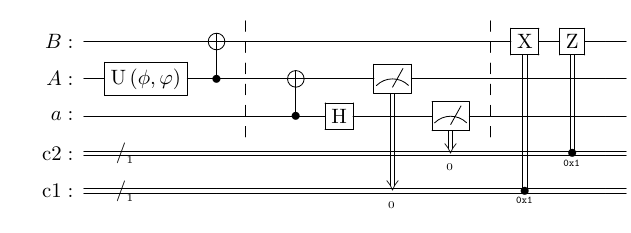}
  \end{center}
  \caption{ Scheme of the standard teleportation protocol. Vertical dashed lines divide the different stages that correspond, from left to right, to: i) Preparation of the initial resource state of the system $A+B$; ii) Application of a Bell measurement in Alice's qubits ($a$ and $A$); iii) Post-processing according to Alice's results.} 
  \label{esquemaObjT1}
\end{figure}
Figure \ref{esquemaObjT1} shows a schematic generalization of the standard teleportation protocol. The qubits $A$ and $B$ are initially in the state $\ket{0}$, whereas $a$ is already in $\rho_{\textrm{in}}$, the state to be teleported. Vertical dashed lines separate the different stages of the protocol corresponding, from left to right, to: 
\begin{itemize}
    \item  A unitary transformation $U(\phi,\varphi)$ rotates the qubit $A$, and a \textsc{cnot} gate is employed to prepare the initial resource state 
    \begin{equation}\label{Irec}
        \ket{\phi_0}_{AB}= \cos\phi \ket{00} + e^{i\varphi}\sin\phi\ket{11},
    \end{equation}
    with $\phi\in[0,\pi/2]$, and $\varphi\in[-\pi/2,3\pi/2]$. (This generalizes the application of the Hadamard gate,  resulting in the Bell state $\frac{1}{\sqrt{2}}(\ket{00}+\ket{11})$).
    
   \item Alice applies a \textsc{cnot} and a Hadamard gate to the qubits in her possession, then performs a measurement in the Bell basis and communicates the outcome to Bob via classical channels.

    \item  Depending on Alice's measurement outcome, Bob applies appropriate unitary operations on his qubit $B$, so the output state of $B$ coincides with the input state of $a$. The whole strategy leads to perfect teleportation if the resource state (\ref{Irec}) is the Bell state $ \frac{1}{\sqrt{2}} (\ket{00}+\ket{11})$ (if  
$\ket{\phi_0}$ was another Bell state, then the operations in Bob's strategy must be changed).  
    
\end{itemize}


For pure input states, $\rho_{\textrm{in}}=\ket{\chi}\!\bra{\chi}$, the success of the teleportation can be quantified by means of the \emph{maximal average fidelity} $F_{\max}$,
which measures the probability that $\rho_{\textrm{out}}$ coincides with the (unknown) $\rho_{\textrm{in}}$, averaged over all input states, provided the appropriate $\sigma^{(i)}$ is chosen. 
The maximal average fidelity for an \emph{arbitrary} 2-qubit resource state $\rho_{AB}$ can be written as \cite{Fsfraction}    
\begin{equation}\label{fmaxdef}
F_{\max}= \frac{1}{3}\Big[2\mathcal{F}_{\max}(\rho_{AB}) +1\Big],
\end{equation}
where $\mathcal{F}_{\max}$ is the \textit{maximal singlet fraction} \cite{Sfranction} 
\begin{equation}\label{maxsingletA}
\mathcal{F}_{\max}= \max_{i}  \{\langle \Phi_i\vert  \rho_{AB}  \vert \Phi_i \rangle \}
\end{equation}
corresponding to the maximum fidelity between the resource state and any of the Bell states 
\footnote{In Ref. \cite{Sfranction}  the maximal singlet fraction is defined considering the maximum over \emph{all} the maximally entangled states. Here we maximize only over those states that can be obtained from Bell states by means of unitary transformations of the form $\mathsf{I}_2\otimes \sigma^{(i)}$. With this restriction we adhere to the standard teleportation protocol (in which Bob's operations are implemented via Pauli operators).}
\beq 
\ket{\Phi_i}=(\mathsf {I}_{2}\otimes\sigma^{(i)})\ket{\Phi^+},\quad \ket{\Phi^+}=\frac{1}{\sqrt 2}(\ket{00}+\ket{11}).
\eeq
Putting $\sigma^{(0)}=\mathsf I_2$, and $\sigma^{(1),(2),(3)}=\sigma^{x,y,z}$, we get 
\beq \label{bellsb}
\begin{split}
\ket{\Phi_{0/3}}= \ket{\Phi^{\pm}}=\frac{1}{\sqrt 2}(\ket{00}\pm\ket{11}), \\
\ket{\Phi_{1/2}}= \ket{\Psi^{\pm}}=\frac{1}{\sqrt 2}(\ket{01}\pm\ket{10}).
\end{split}
\eeq
Therefore, $\mathcal F_{\max}$ picks out the optimal Bell state, i.e., the Bell state that is closest to $\rho_{AB}$. 
The above expressions show that if $\rho_{AB}$ is a Bell state, the protocol guarantees 
the complete reconstruction of the input state. 

For the quantum teleportation to be considered successful  $F_{\max}$ must be greater than $2/3$, which is the value of the maximal average fidelity corresponding to the best possible reconstruction through a purely classical channel \cite{Fclassical}. Throughout this paper we will be interested in the success of non-classical teleportation, hence we will focus on those regimes in which $F_{\max}\geq 2/3$, and accordingly pay attention to the quantity
\beq \label{fmaxnos}
\mathbb F_{\max}=\max \Big\{\frac{2}{3},F_{\max}\Big\}.
\eeq


 \section{Maximal average fidelity under quantum channels} \label{FK}

\begin{figure}[ht]
  \begin{center}
    \includegraphics[scale=1]{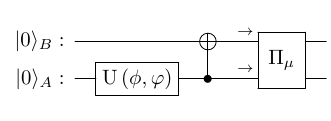}
  \end{center}
  \caption{The pair of qubits $A$ and $B$, once in the entangled state $\ket{\phi_{0}}$, passes through an arbitrary quantum channel represented by the set of Kraus operators $\Pi_{\mu}$, resulting in a mixed resource state $\rho_{AB}$. } 
  \label{esquemaObjT11}
\end{figure}

Let us assume that $A$ and $B$ are initially prepared in the state $\ket{\phi_0}_{AB}$, given by Eq. (\ref{Irec}), which then passes through a quantum channel $\Lambda_{AB}$
represented by a set of Kraus operators  $\{\Pi_{\mu}\}$, acting on $\mathcal{H}_A\otimes \mathcal{H}_B$ \cite{Kraus}. A scheme of this process is shown in Fig. \ref{esquemaObjT11}.
The channel thus transforms $\ket{\phi_0}\!\bra{\phi_0}$ into an effective (typically mixed) resource state $\rho_{AB}$ given by 
\begin{equation}\label{resourcegeneral}
\rho_{AB}=\Lambda_{AB}(\ket{\phi_{0}}\!\bra{\phi_{0}})=\sum_{\mu} \Pi_{\mu} \ket{\phi_{0}}\!\bra{\phi_{0}}  \Pi^{\dagger}_{\mu}.
\end{equation}

\noindent Direct substitution  into  Eq.~(\ref{maxsingletA}) gives, with the aid of (\ref{fmaxdef}),
\begin{equation}\begin{split}\label{fmaxKB}
F_{\max}&=\frac{1}{3} + \frac{2}{3} \max_{i}  \Big\{ \sum_{\mu} \big\vert \langle\Phi_i\big\vert \Pi_{\mu} \big\vert\phi_{0}\rangle 
\big\vert^{2} \Big\}.
\end{split}\end{equation}
This simple but central expression for $F_{\max}$ in terms of the Kraus operators allows for studying the teleportation success when the initially pure resource state is subject to an arbitrary quantum channel.   

Since Alice and Bob are typically spatially separated, we will focus on cases in which $A$ and $B$ undergo independent local channels, so  $\Lambda_{AB}= \Lambda_{A}\otimes\Lambda_{B}$ and 
\beq\label{Krauslocal}
\Pi_{\mu}\rightarrow\Pi_{\alpha\beta}=Q_\alpha\otimes K_\beta,
\eeq
where $\{Q_\alpha\}$ and $\{K_\beta\}$ stand for the sets of Kraus operators associated to $\Lambda_A$ and $\Lambda_B$, respectively, each set having at most $(\dim\mathcal{H}_{A(B)})^2=4$ elements.
Equation (\ref{fmaxKB}) becomes then
\begin{equation}\begin{split}\label{fmaxKBlocal}
F_{\max}&=\frac{1}{3} + \frac{2}{3} \max_{i}  \Big\{ \sum_{\alpha\beta} \big\vert \langle\Phi_i\big\vert Q_{\alpha}\otimes K_{\beta} \big\vert\phi_{0}\rangle  \big\vert^{2} \Big\}.
\end{split}\end{equation} 

The channel $\Lambda_A\otimes \Lambda_B$
acts as if $A$ and $B$ interact locally with a corresponding party $E_A$ and $E_B$. Assuming that the systems $(A+B)$, $E_A$ and $E_B$ are 
initially uncorrelated, and that $E_{A}$ and $E_B$ are qubits in the initial state $\ket{0}$, we may write
\beq\label{initAB_EAEB}
\ket{\psi_{0}}_{ABE_AE_B}=\ket{\phi_0}_{AB}\otimes\ket{0}_{E_A}\otimes\ket{0}_{E_B}
\eeq
for the initial 4-partite state. Further, if the interaction between $A$ and $E_{A}$ is represented by the unitary operator $U_{AE_A}$ (and similarly for $B$ and $E_B$), then 
\beq
\begin{split}\label{Kraus1}
Q_{\alpha}=  {}_{E_A}\!\bra{\alpha} U_{AE_A}\ket{0}_{E_A},\\ K_{\beta}= {}_{E_B}\!\bra{\beta} U_{BE_B}\ket{0}_{E_B},
\end{split}
\eeq
with $\{\ket{\alpha}\}$ and $\{\ket{\beta}\}$ basis of $\mathcal H_{E_A}$ and $\mathcal H_{E_B}$, respectively. 
In addition, the (unitary) evolution of the complete system can be obtained from these Kraus operators as
\begin{eqnarray}
\label{evolstate}
\ket{\psi}_{ABE_AE_B}&=&U_{AE_A}U_{BE_B}\ket{\psi_0}_{ABE_AE_B}\nonumber\\
&=&\sum_{\alpha \beta}Q_{\alpha}K_{\beta}\ket{\phi_{0}}_{AB}\ket{\alpha}_{E_A}\ket{\beta}_{E_B}.
\end{eqnarray}

A particular family of local channels, that involves a 3-party system instead of a 4-partite one, is that in which one of the resource qubits, say $A$, remains unaffected so $\Lambda_{A}=\mathsf{I}_2$, while $B$ goes through an arbitrary channel  $\Lambda_{B}$.  In this case $Q_\alpha=\mathsf I_2\, \delta_{\alpha 0}$, and from Eq. (\ref{fmaxKBlocal}) we are led to
\beq\label{fmaxrecexplicit}
F_{\max}= \frac{1}{3} + \frac{2}{3} \max_{i}  \Big\{ \sum_{\beta} \big\vert \langle\Phi_i\big\vert (\mathsf I_2\otimes K_{\beta})\vert\phi_{0}\rangle  \big\vert^{2} \Big\}.
\eeq
Clearly for $\Lambda_{AB}=\mathsf I_2\otimes \Lambda_B$ the subsystem $E_A$ is superfluous, and Eq. (\ref{initAB_EAEB}) reduces to the 3-qubit initial state
\beq\label{initAB_EB}
\ket{\psi_{0}}_{ABE_B}=\ket{\phi_0}_{AB}\otimes\ket{0}_{E_B}.
\eeq

\section{Fidelity and bipartite entanglement in the standard teleportation protocol}\label{bipartite}

The standard protocol, depicted in the circuit of Fig. \ref{esquemaObjT1}, 
corresponds to $\Lambda_{AB}=\mathsf{I}_4$. Equation (\ref{fmaxKB}) thus becomes 
\beq\label{fmaxnointeractionA}
F_{\max}=\frac{1}{3} + \frac{2}{3} \max_{i}  \Big\{  \big\vert \langle\Phi_i\big\vert \phi_{0}\rangle  \big\vert^{2} \Big\}.
\eeq
With $\ket{\phi_0}$ given by (\ref{Irec}), the optimal strategy corresponds to the state  $\ket{\Phi^{+}}$  for $\varphi\in[-\frac{\pi}{2},\frac{\pi}{2}]$, and $\ket{\Phi^{-}}$ for $\varphi\in[\frac{\pi}{2},\frac{3\pi}{2}]$, so (\ref{fmaxnointeractionA}) reduces to $F_{\max}=\frac{2}{3}+\frac{1}{3}\,\mathcal E_0 \,|\!\cos\varphi|$, and consequently, in the non-interacting case, Eq. (\ref{fmaxnos}) gives
\beq
\label{fmaxnointeraction}
\mathbb F^{\textrm{non-int}}_{\max}=\frac{2}{3}+\frac{1}{3}\,\mathcal E_0 \,|\!\cos\varphi|,
\eeq
where $\mathcal E_0$ stands for the entanglement of the initial resource state  
\begin{equation}
\label{initialtangle}
\mathcal{E}_{0}\equiv C(\vert\phi_{0}\rangle)= 2\cos\phi\sin\phi=\sin 2\phi.
\end{equation}
Here $C$ is the \emph{concurrence}, quantifying the amount of qubit-qubit entanglement \cite{Concurrencia}. 
For an arbitrary (in general mixed) 2-qubit state $\rho_{AB}$ it is defined as
\begin{equation}\label{concumixed}
C_{AB}\equiv C(\rho_{AB})= \max\{0,\sqrt{\lambda_{0}}-\sqrt{\lambda_{1}}-\sqrt{\lambda_{2}}-\sqrt{\lambda_{3}}\},
\end{equation}
where $\{\lambda_{n}\}$ are the eigenvalues of the  matrix $\rho_{AB}(\sigma^{y}\otimes\sigma^{y})\rho^{*}_{AB}(\sigma^{y}\otimes\sigma^{y})$
ordered in decreasing order, and $\rho^{*}_{AB}$ stands for the complex conjugate of $\rho_{AB}$ expressed in the computational basis. 
When the state is pure ($\rho_{AB}=\ket{\chi}\!\bra{\chi}$), the expression for the concurrence simplifies and reads
\begin{equation}\label{concupure}
C_{AB}=C(\ket{\chi})=\sqrt{2(1-\textrm{Tr}\rho^2_A)}=\sqrt{2(1-\textrm{Tr}\rho^2_B)},
\end{equation}
with $\rho_{A(B)}=\textrm{Tr}_{B(A)}\rho_{AB}$ the reduced density matrix of either one of the qubits.

\begin{figure}[ht]
  \centering
    \includegraphics[scale=0.8]{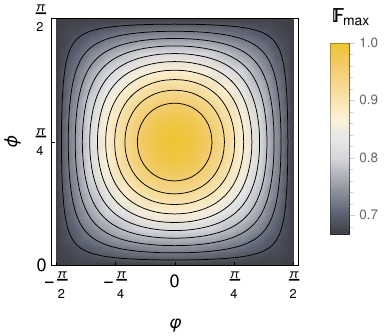}
  \caption{Maximal average fidelity in absence of interaction (Eq. (\ref{fmaxnointeraction})) as a function of $\varphi$ and $\phi$, in color scale ranging from $2/3$ to 1. At $\varphi=\pm\pi/2$, $\mathbb F^{\textrm{non-int}}_{\max}$ shows no improvement with respect to the classically attainable value ($2/3$), irrespective of $\phi$, which determines the initial entanglement $\mathcal E_0=\sin2\phi$.}
  \label{nointeraction}
\end{figure}

Equation (\ref{fmaxnointeraction}) makes explicit that the teleportation success enhances as the resource state's entanglement $\mathcal E_0$ increases. 
It also shows that as $\varphi$ tends to $\pm\pi /2$, the fidelity decreases up to its minimal ---classically attainable--- value $2/3$, irrespective of the initial entanglement, as can be seen in Fig. \ref{nointeraction}.
That is, there are maximally entangled resource states for which the maximal average fidelity does not exceed its classical limit, and the relative phase $\varphi$ determines the fraction of $\mathcal E_0$ that ultimately improves the fidelity. 


\section{Fidelity and three-partite entanglement generated via a noisy channel}\label{Stripartite}

\begin{figure}[ht]
  \begin{center}
\includegraphics[scale=0.9]{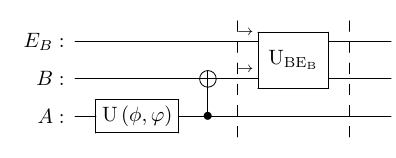}
  \end{center}
  \caption{Qubits $A$, $B$ and $E_B$ are prepared in the initial state (\ref{initAB_EB}). Then $B$ and $E_B$ interact via a unitary operation $U_{BE_B}$, and the effect on $B$ is that of a quantum channel described by the Kraus operators $\{K_\beta\}$.  }
  \label{esquemaObjT3}
\end{figure}

We now focus on the case in which only $B$ undergoes through a quantum channel, effectively representing an interaction with an additional qubit $E_B$. Figure \ref{esquemaObjT3} illustrates this situation, leading to a mixed resource state $\rho_{AB}$ before the measurement stage of the protocol. 

The scenario under consideration corresponds to that in which $\ket{\phi_0}\!\bra{\phi_0}$ is subject to a channel $\Lambda_{AB}=\mathsf I_2\otimes \Lambda_B$, where $\Lambda_B$ encodes the interaction between $B$ and $E_B$, giving rise to the possible creation of tripartite entanglement in the system $A+B+E_B$. 
The channel thus transforms the initial state (\ref{initAB_EB})
\beq
\label{initABEbis}
\ket{\psi_{0}}_{ABE_B}=\cos\phi \ket{000}+  e^{i\varphi}\sin\phi \ket{110}
\eeq
into the 3-qubit state (see Eq. (\ref{evolstate}) with $Q_\alpha=\mathsf I_2 \,\delta_{\alpha 0}$)
\beq
\label{evolstate3q}
\ket{\psi}_{ABE_B}=\sum_{ \beta}K_{\beta}\ket{\phi_{0}}_{AB}\ket{\beta}_{E_B}.
\eeq

In \cite{nuestro} necessary and sufficient conditions on the Kraus operators $K_{\beta}=\bra{\beta} U_{BE_B} \ket{0}$ were established, that ensure the emergence of bipartite and tripartite entanglement among the parties $A,B,E_B$.
For the present analysis we concentrate on the dynamics of the resource state's (bipartite) entanglement  $C_{AB}=C(\rho_{AB})$, and the tripartite entanglement, as measured by the so-called \emph{3-tangle}. The latter stands as a legitimate measure of residual entanglement in a 3-qubit pure state $\ket{\psi}_{ijk}$, and quantifies the amount of three-way \textsc{ghz}-type entanglement in the state \cite{DurPRA2000}. The 3-tangle is defined as \cite{ckw}
\beq \label{tau_def}
\tau_{ijk}=\tau(\ket{\psi}_{ijk})=C^{2}_{i|jk}-C^{2}_{ij}-C^{2}_{ik},
\eeq
where $C_{ij}$ is given by Eq. (\ref{concumixed}), and $C_{i|jk}$ by
\beq
C_{i|jk}=\sqrt{2(1-\textrm{Tr}\rho^2_i)}.
\eeq
The last expression generalizes (\ref{concupure}) for a 3-party pure state and quantifies the entanglement across the bipartition $i|(j+k)$ \cite{RungtaPRA01}. 

In \cite{nuestro} it is found that for $\Lambda_{AB}=\mathsf I_2\otimes \Lambda_B$, $C_{AB}$ and $\tau_{ABE_B}$ evolve according to
\begin{eqnarray}
\label{resourceconcurence3}
C^{2}_{AB}&=& \mathcal{E}^{2}_{0}(\vert\det K_{0}\vert +\vert\det K_{1}\vert)^{2}\nonumber\\
&&-\frac{1}{2}\mathcal{E}^{2}_{0}\big(\vert u \vert - \vert v\vert +\vert v-u\vert \big),
\end{eqnarray}
and
\beq
\label{tau3}
\tau_{ABE_B}=\mathcal{E}^{2}_{0}\vert u-v \vert,
\eeq
where $u=4\det K_{0}K_{1}$ and $v=g^{2}(K_{0},K_{1})$, with $g(M,N)=\textrm{Tr} M \,\textrm{Tr} N - \textrm{Tr} (MN)$. 
The evolution parameter is encoded in the Kraus operators and is not explicitly written in the above expressions.

\subsection{Generalized noisy channel}

In order to study the role of the 3-tangle in the teleportation fidelity, we focus on   
channels $\Lambda_B$ whose Kraus operators have the following structure \cite{RuanPRA2018}  (in the basis $\{\ket{0}=(1,0)^{\top},\ket{1}=(0,1)^{\top}\}$ of $\mathcal{H}_B$)
%
\begin{equation}\label{krausACDC}
K_{0}=\left(\begin{array}{cc} 1 & 0 \\ 0 & \sqrt{1-p} \end{array}\right), \quad K_{1}= \sqrt{p} \left(\begin{array}{cc} 0 & \cos\zeta \\ 0 & \sin\zeta \end{array}\right),
\end{equation}
where $\zeta\in [0,\pi/2]$ and $0\leq p\leq 1$. From the second equation in (\ref{Kraus1}) and the explicit form of $K_0$ in (\ref{krausACDC}), we get $\bra{1} K_{0}\ket{1}=\bra{10} U_{BE_B}\ket{10}=\sqrt{1-p}$. This means that under $U_{BE_B}$, the state $\ket{10}$ transforms into a state that writes as 
\beq \label{prob}
U_{BE_B}\ket{10}=\sqrt{1-p}\ket{10}+\sqrt{p}e^{i\theta}\ket{10_{\perp}},
\eeq
with $\ket{10_{\perp}}$ a normalized state that is orthogonal to $\ket{10}$. Consequently, $p$ can be interpreted as the probability that the state $\ket{10}_{BE_B}$ evolves into an orthogonal state under the transformation $U_{BE_B}$. Clearly $p$ is a function of the evolution parameter of $U_{BE_B}$, typically the time for Hamiltonian evolutions $U_{BE_B}(t)=e^{-iHt/\hbar}$. 
Initially (for $U_{BE_B}(0)=\mathsf I_4$) $p$ vanishes, and increases up to $p=1$ when the  state $\ket{10_{\perp}}$ (completely distinguishable from $\ket{10}$) is reached. This allows us to identify $p$ as a useful parameter to track the evolution induced by $U_{BE_B}$, without making specific assumptions regarding such unitary transformation. 

Each value of the parameter $\zeta$ in (\ref{krausACDC}) determines a specific channel, so comparison of the dynamics under different channels can be achieved by varying the values of $\zeta$. When $\zeta=0$, (\ref{krausACDC}) reduce to the Kraus operators of the \textit{amplitude damping channel} (\textsc{ac}), whereas for $\zeta=\pi/2$ the Kraus operators of the  \textit{dephasing channel} (\textsc{dc}) are recovered. 
The \textsc{ac} and the \textsc{dc} are paradigmatic decoherence channels \cite{Aolita_2015} that generate, respectively, \textsc{w}-type and \textsc{ghz}-type genuine entanglement in the 3-qubit system \cite{nuestro,FariasPRL1092012,AguilarPRA892014}. 
By means of Eq. (\ref{krausACDC}) ---corresponding to what we will call the \emph{generalized (noisy)} channel, or \textsc{gc} for short--- we can analyze intermediate situations lying between the \textsc{ac} and the \textsc{dc}, and particularly extend some of the results reported in, e.g., \cite{Laura2014,IshizakaPRA2001,VanhopQIP2019}, to a wider range of channels.

When $B$ is subject to the \textsc{gc}, the evolved state (\ref{evolstate3q}) reads explicitly 
\begin{equation}\label{psiACDC}
\begin{split}
\vert\psi\rangle=& \cos\phi\ket{000} + e^{i\varphi}\sin\phi  \big( \sqrt{1-p}\ket{110} + \\ 
&+ \sqrt{p}\cos\zeta \ket{101} + \sqrt{p}\sin\zeta\ket{111} \big).
\end{split}
\end{equation}
The entanglement of the corresponding resource state and the generated amount of 3-tangle become, using Eqs. (\ref{resourceconcurence3}) and (\ref{tau3}), 
\begin{equation}\label{concuACDC}
C_{AB}=\mathcal{E}_{0}\sqrt{1-p},\quad \tau_{ABE_B}= \mathcal{E}^{2}_{0}\,p \sin^{2}\zeta,
\end{equation}
so only the 3-tangle depends on the specific channel. 
Further, $p \sin^{2}\zeta$ determines the fraction of the initial entanglement $\mathcal E_0$ that can be converted into 3-partite entanglement.
For fixed $\mathcal E_0$ and $p$, as $\zeta$ increases from $0$ to $\pi/2$ the 3-tangle goes from its minimum ($0$) to its maximum ($\mathcal E^2_0 \,p$) value. 
The minimum corresponds to the \textsc{ac} case ($\zeta=0$), and the maximum to the \textsc{dc} case ($\zeta=\pi/2$), which is the only channel for which all the initial entanglement can be transformed into 3-tangle (at $p=1$).

From Eq. (\ref{fmaxrecexplicit}) it follows that for $B$ subject to the \textsc{gc}, the maximal average fidelity can be written as 
$F_{\max}=\max_{\,i}\,\{F_{\Phi_{i}}\}$,
with
\beq
F_{\Phi_{i}}=\frac{1}{3} + \frac{2}{3} \sum_{\beta} \big\vert \langle\Phi_i\big\vert (\mathsf I_2\otimes K_{\beta})\vert\phi_{0}\rangle  \big\vert^{2}. 
\eeq
Direct calculation gives
\begin{subequations}
\begin{eqnarray}
\label{casosF_uncanal}
F_{\Phi^{\pm}}\!\!&=& \!\frac{2}{3} +\frac{1}{3}\Big(\! \pm\mathcal{E}_{0} \sqrt{1-p}\cos\varphi- \mathcal P_1 p\cos^{2}\zeta\Big),  \\
F_{\Psi^+}\!\!&=&\!F_{\Psi^{-}}= \frac{2}{3}  -\frac{1}{3}\Big(1-\mathcal P_1 p\cos^{2}\zeta\Big),
\end{eqnarray}
\end{subequations}
where
\beq\label{pop1}
\mathcal P_1=\sin^2\!\phi
\eeq
stands for the initial population of the state $\ket{11}_{AB}$. 
On one hand  $\mathcal{P}_{1}p\cos^{2}\zeta \leq 1$ implies that $F_{\Psi^{\pm}}\leq 2/3$. 
On the other hand,
\begin{eqnarray}\label{maxmasmenos}
\max\{F_{\Phi^{+}},F_{\Phi^{-}} \}&=&
\begin{cases}
  F_{\Phi^{+}}  & \varphi\in[-\pi/2,\pi/2] \\
  F_{\Phi^{-}} & \varphi\in[\pi/2,3\pi/2]
  \end{cases}\\
&=&
\frac{2}{3} +\frac{1}{3}\Big(\mathcal{E}_{0} \sqrt{1-p}\,|\!\cos\varphi|- \mathcal P_1 p\cos^{2}\zeta \Big).\nonumber
\end{eqnarray}
Consequently, 
under the generalized channel, the quantity of interest (\ref{fmaxnos}) is given by
\beq \label{FmaxACDC}
\mathbb F^{\textsc{gc}}_{\max}=\frac{2}{3}+\frac{1}{3}\max\Big\{0,\Big[C_{AB}(p)\,|\!\cos\varphi|- \mathcal P_1 p\cos^{2}\zeta\Big]\Big\},
\eeq
where we have used Eq. (\ref{concuACDC}) for $C_{AB}$.
\begin{figure*}
\includegraphics[scale=0.65]{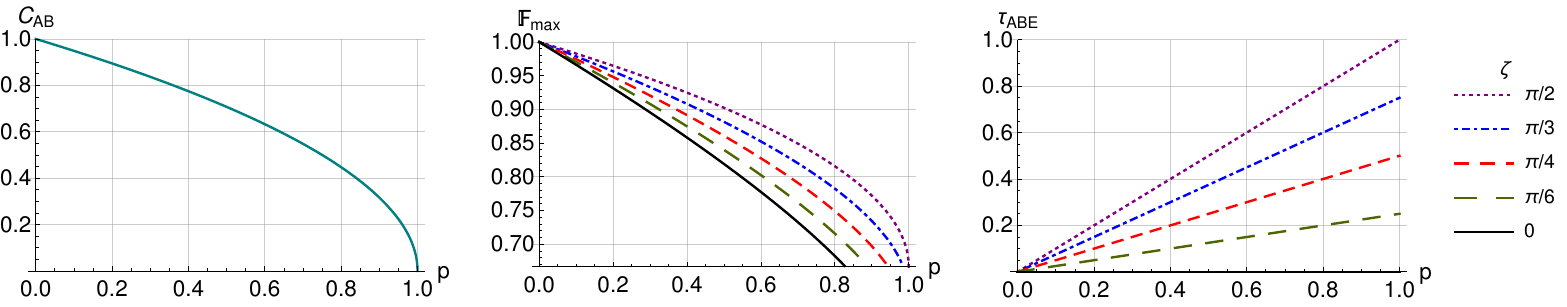}
  \caption{Evolution of the entanglement of the resource state (left panel), the maximal average fidelity above the classical threshold value (central panel), and the 3-tangle (right panel) when the qubit $B$ is subject to a \textsc{gc} channel, for different values of the channel parameter $\zeta\in [0,\pi/2]$, and  $\phi=\pi/4$, $\varphi=0$. The channel induces a loss in the resource entanglement 
$C_{AB}$, and the teleportation fidelity decreases accordingly. However, when different channels (values of $\zeta$) are considered, it is observed that those that generate more 3-tangle favor higher fidelities at each $p$. Therefore, as $\zeta$ increases from $0$ (amplitude damping channel) to $\pi/2$  (dephasing channel), the generated 3-tangle helps to enhance the maximal average fidelity (provided $C_{AB}\neq0)$.}
\label{f3qubitevol}
\end{figure*}
It follows from Eq. (\ref{FmaxACDC}) that, as in the noiseless scenario (see Eq. (\ref{fmaxnointeraction})), when the maximal average fidelity exceeds the classical threshold it has a non-negative contribution proportional to the entanglement of the resource state, attenuated by $|\!\cos\varphi|$. 
This indicates that, as in the noiseless case, the optimal $\varphi$ is $0,\pi$. 
In the \textsc{gc} case, however, an additional negative contribution (a loss in the fidelity) appears that depends on the channel (via $\cos^{2}\zeta$), the evolution ($p$), and the initial excited population ($\mathcal P_1$). 
For fixed $\mathcal P_1$  and $p$, 
such loss decreases as $\tau_{ABE_B}\sim \sin^2\zeta$ increases. 
This means that given an initial state, and \emph{at each stage of the evolution} (determined by a fixed $p$), the channels that produce higher amounts of 3-tangle lead to higher values of the maximum average fidelity. 

The above conclusion can also be extracted from Figure \ref{f3qubitevol}, which shows $C_{AB}$, $\mathbb F^{\textsc{gc}}_{\max}$, and $\tau_{ABE_B}$ for different values of $\zeta$ and varying $p$, considering a maximally entangled initial state ($\phi=\pi/4$) and setting $\varphi$ to its optimal value $\varphi=0$.
At $p=0$, $\mathbb F^{\textsc{gc}}_{\max}$ reduces to $\mathbb{F}^{\textrm{non-int}}_{\max}$, no 3-tangle exists, and the success of the teleportation is ascribable to the (bipartite) entanglement $\mathcal E_0=C_{AB}(0)$ only, in line with the discussion in Sec. \ref{bipartite}.
As $p$ increases, the action of $\Lambda_B$ degrades the entanglement of the resource state (whose dynamics is independent of $\zeta$), 
and a concomitant gradual loss in the maximal average fidelity is observed for all $\zeta$. 
This is an expected behavior under generic local  channels $\Lambda_A\otimes\Lambda_B$ (meaning non-increasing entanglement operations). 

In its turn, as $p$ increases the 3-tangle increases as well, as a result of the redistribution of the entanglement. 
Therefore, for any given generalized noisy channel, the decay of the maximal average fidelity along the evolution is accompanied by an increase of the 3-tangle. 
The question then arises as to under which channels  $\mathbb F^{\textsc{gc}}_{\max}$ is more robust against the noise, that is, under which channels the loss in the teleportation success is reduced at any stage of the evolution. 
To answer it, we compare the values of $\mathbb F^{\textsc{gc}}_{\max}$ for all possible \textsc{gc} channels while keeping $p$ fixed. 
Comparison of the central and rightmost panels in Fig. \ref{f3qubitevol} shows that the decaying path of $\mathbb F^{\textsc{gc}}_{\max}$ varies with $\zeta$ in such a way that for each and every $p$, the channels that generate higher amounts of 3-tangle yield higher fidelities. 
Table \ref{table3qubits} exemplifies this by explicitly showing the values of $\mathbb F^{\textsc{gc}}_{\max}$ and $\tau_{ABE_B}$ extracted from Fig. \ref{f3qubitevol} for $p=0.8$. While the entanglement of the resource state is constant as $\zeta$ varies (see the first equation in (\ref{concuACDC})), the maximal average fidelity and the 3-tangle do change in a correlated fashion: an increase in $\tau_{ABE_B}$ is accompanied by an increase in $\mathbb F^{\textsc{gc}}_{\max}$. This behavior replicates for any other value of $p\in(0,1)$.
%
\begin{table}[h]
    \centering
    \begin{tabular}{|c|c|c|c|}
    \hline
   $\zeta$& $C_{AB}$ &$\mathbb{F}^{\textsc{gc}}_{\max}$ & $\tau_{ABE_B}$
    \\ 
     \hline
     \hline
    0&  0.447214& 0.682405 & 0 
     \\ \hline
    $\pi/6$ &  0.447214 & 0.715738 & 0.2
    \\ \hline
    $\pi/4$ &  0.447214&  0.749071 & 0.4 
    \\ \hline
    $\pi/3$& 0.447214& 0.782405 & 0.6 
    \\ \hline
    $\pi/2$ &  0.447214& 0.815738 & 0.8 
    \\ \hline
    \end{tabular}
    \caption{Values of $C_{AB}$, $\mathbb{F}^{\textsc{gc}}_{\max}$, and $\tau_{ABE_B}$ taken from Fig. \ref{f3qubitevol} (corresponding to $\phi=\pi/4$ and $\varphi=0$) for $p=0.8$ and different values of the channel parameter $\zeta$.}
    \label{table3qubits}
\end{table}

Therefore, despite the adverse influence of noise on the fidelity, the detrimental effects are lessened under channels that give rise to higher amounts of tripartite entanglement. 
It is in this sense that the 3-tangle improves the teleportation success, and may help to maintain $\mathbb F^{\textsc{gc}}_{\max}$ above the classical threshold value longer (i.e., for larger values of $p$, as seen in the central panel of Fig. \ref{f3qubitevol}).

It is also clear from Fig. \ref{f3qubitevol} that the channel that gives better fidelities throughout the evolution is the \textsc{dc}, corresponding to $\zeta=\pi/2$. 
Only for this channel $\mathbb F^{\textsc{gc}}_{\max}$ exceeds the classically attainable value for all $0<p<1$. 
This holds not only for a maximally entangled initial state, but also for $\mathcal E_0\neq 0$. 
In fact, it follows from Eq. (\ref{FmaxACDC}) that the condition $\mathbb F^{\textsc{gc}}_{\max}>2/3$ amounts to 
\beq\label{1blancas}
\mathcal P_1p\cos^2\zeta<C_{AB}(p) |\!\cos\varphi|=\mathcal E_0\sqrt{1-p}\,|\!\cos\varphi|,
\eeq
which is trivially satisfied for $p\in(0,1)$ and $\mathcal E_0\neq 0$ taking $\zeta=\pi/2$. 
Further, by writing $\mathcal E_0=2\sqrt{\mathcal P_1(1-\mathcal P_1)}$, the condition (\ref{1blancas}) rewrites as
\beq\label{blancas}
\frac{1}{2}\frac{\cos^2\zeta}{|\!\cos\varphi|}\frac{p}{\sqrt{1-p}}<\sqrt{\frac{1-\mathcal P_1}{\mathcal P_1}},
\eeq
which exhibits  the role of the initial population $\mathcal P_1$ in the dynamics of $\mathbb F^{\textsc{gc}}_{\max}$ (assuming fixed $\varphi$ and $\zeta$):
the left-hand side of the inequality is an increasing  function of $p$, whereas the upper bound decreases with $\mathcal P_1$; 
consequently, as $\mathcal P_1$ increases the inequality becomes more restrictive, and a point in the evolution is  reached sooner (for lower values of $p$) at which   
(\ref{blancas}) does not longer hold, and the fidelity drops below the threshold value $2/3$.  
\begin{figure*}
\includegraphics[scale=0.65]{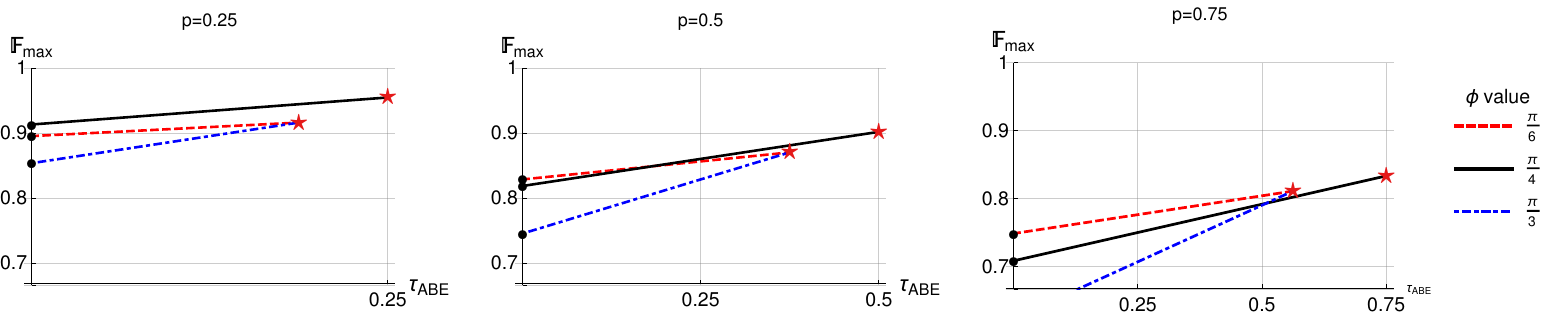}
\caption{Trajectories in the space $(\tau_{ABE_B},\mathbb F^{\textsc{gc}}_{\max})$ as the channel parameter $\zeta$ runs from $0$ (black circles) to $\pi/2$ (red stars), for three initial states with  $\phi=\pi/6$ (red dashed line), $\phi=\pi/4$ (black solid line), and $\phi=\pi/3$ (blue dotted-dashed line), at three stages of the evolution: $p=0.25$ (left), $p=0.5$ (center), and $p=0.75$ (right). In all cases the relative phase was fixed to its optimal value $\varphi=0$.
For each initial state, and at 
fixed $p$, the channels that produce higher 3-tangle lead to higher maximal average fidelities. A non-trivial role of the initial entanglement $\mathcal E_0$ and the initial excited population $\mathcal P_1$ is observed (see text).}
  \label{c3qubit}
\end{figure*}

Finally, as $p\rightarrow1$ we have $C_{AB}\rightarrow0$, and from Eq. (\ref{FmaxACDC}) $\mathbb F^{\textsc{gc}}_{\max}$ cannot exceed $2/3$ despite $\tau_{ABE_B}$ attains its maximum value. This evinces that the 
even though the presence of tripartite entanglement improves the teleportation success (in the sense described above), a non-zero entanglement of the resource state is key to trigger the potential of the 3-tangle to assist the teleportation. 


Figure \ref{c3qubit} shows, at different stages of the evolution, 
curves in the space 
$(\tau_{ABE_B},\mathbb F^{\textsc{gc}}_{\max})$ %
sweeping as the channel parameter goes from $\zeta=0$ (\textsc{ac}, black circles) to $\zeta=\pi/2$ (\textsc{dc}, red stars). 
Each trajectory corresponds to an initial state determined by  $\phi=\pi/6$ (red dashed line), $\phi=\pi/4$ (black solid line), and $\phi=\pi/3$ (blue dotted-dashed line), all with $\varphi=0$.
In all curves the relation between the 3-tangle and the maximal average fidelity discussed above is manifest:
throughout the evolution,  better fidelities are attained under channels that produce higher amounts of 3-tangle.

Interestingly,  Fig. \ref{c3qubit} shows that whereas an initial resource state with maximal entanglement  ($\phi=\pi/4$) leads to the highest value of $\mathbb F^{\textsc{gc}}_{\max}$ (for sufficiently large values of $\tau_{ABE_B}$), a maximally entangled state is not always the optimal one for achieving a better fidelity. 
This is clearly seen, for example, in the rightmost panel of Fig. \ref{c3qubit}, where the fidelity for $\phi=\pi/6$ ($\mathcal E_0\sim 0.866$) is greater than that corresponding to $\phi=\pi/4$ ($\mathcal E_0=1$) for some channels. 
Moreover, noticing that $\phi=\pi/3$ gives the same $\mathcal E_0$ as $\phi=\pi/6$, the difference between the red dashed and the blue dotted-dashed curves brings out the effect of the initial population $\mathcal P_1$ on the behavior of $\mathbb F^{\textsc{gc}}_{\max}$. 
In particular, for $\phi=\pi/3$ we have $\mathcal P_1=0.75$, while for $\phi=\pi/6$ we have $\mathcal P_1=0.25$. 
As seen from Eq. (\ref{FmaxACDC}), the greater $\mathcal P_1$, the greater the loss in the fidelity (which goes as $-\mathcal P_1p\cos^2\zeta$), which explains why the blue dotted-dashed line runs below the red dashed one, despite both curves correspond to the same initial entanglement.


In summary, from all the generalized channels acting on $B$, those that improve the quantum teleportation success (at each instant) are the ones that produce higher amounts of 3-tangle among the qubits $A, B, E_B$. 
The optimal channel corresponds to $\zeta=\pi/2$ (\textsc{dc}), and the worst to $\zeta=0$ (\textsc{ac}). 
This can be seen graphically, and is corroborated by Eq. (\ref{FmaxACDC}), which gives
\beq
\mathbb{F}^{\textsc{ac}}_{\max}\leq \mathbb{F}^{\textsc{gc}}_{\max}\leq \mathbb{F}^{\textsc{dc}}_{\max},
\eeq
with
\beq \label{FmaxDC}
\mathbb{F}^{\textsc{dc}}_{\max}
=\frac{2}{3} +\frac{1}{3}\,C_{AB}(p)\,|\!\cos\varphi|,
\eeq
and
\begin{eqnarray}\label{FmaxAC}
\mathbb F^{\textsc{ac}}_{\max}
&=&\frac{2}{3}+\frac{1}{3}\max\Big\{0,C_{AB}(p)\,|\!\cos\varphi|- \mathcal P_1p\Big\}\nonumber\\
&=&\max\Big\{\frac{2}{3},\mathbb F^{\textsc{dc}}_{\max}-\frac{1}{3}\mathcal P_1 p\Big\}.
\end{eqnarray}
%
Below we explore in more detail the states produced by the limiting cases  \textsc{dc} and \textsc{ac}.

\subsection{Amplitude Damping Channel vs Dephasing Channel}

The \textsc{dc} produces \textsc{ghz}-type states, whose form follows from Eq. (\ref{psiACDC}) with $\zeta=\pi/2$:
\begin{eqnarray}\label{GHZT}
\ket{\psi_{\textsc{ghz}}}& \equiv&\ket{\psi(\zeta=\pi/2)}\nonumber\\
&=&\cos\phi\ket{000} + e^{i\varphi}\sin\phi  \big( \sqrt{1-p}\ket{110} + \\ 
&&+  \sqrt{p}\ket{111} \big).\nonumber
\end{eqnarray}
If $\phi=\pi/4$ and $\varphi=0$ then, at 
$p=1$, the usual \textsc{ghz} state $\ket{\textsc{ghz}}=\frac{1}{\sqrt 2}\big(\!\ket{000} +    \ket{111}\!\big)$
is reached, having null qubit-qubit entanglement and $\tau_{ABE_B}=1.$

For $\zeta=0$ the \textsc{gc} channel reduces to the \textsc{ac}, and states that are equivalent (up to local unitary transformations) to \textsc{w}-type states arise. This can be seen by putting $\zeta=0$ in Eq. (\ref{psiACDC}), obtaining 
\begin{eqnarray}\label{WT}
\ket{\psi_\textsc{w}}&\equiv&\ket{\psi(\zeta=0)} \nonumber\\
&=&\cos\phi\ket{000} + e^{i\varphi}\sin\phi  \big( \sqrt{1-p}\ket{110} + \\ 
&&+ \sqrt{p} \ket{101} \big).\nonumber
\end{eqnarray}
For $\phi=\arccos(1/\sqrt{3})$ and $\varphi=0$ this state becomes, at $p=1/2$, equivalent (up to a local unitary transformation) to the usual \textsc{w} state $\ket{\textsc{w}}=\frac{1}{\sqrt{3}}\big(\ket{000}+\ket{110} +\ket{101} \big)$,
characterized by having all qubit-qubit entanglements equal to $C_{ij}= 2/3$, and null 3-tangle.

\begin{figure}[ht]
  \centering
    \includegraphics[scale=0.6]{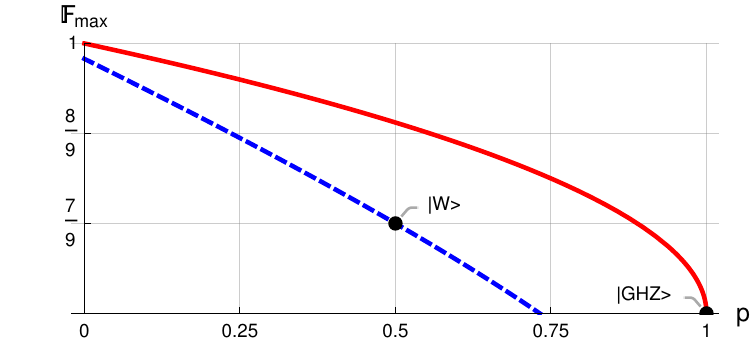}
  \caption{Evolution of the maximal average fidelity (above the threshold value $2/3$), employing initial states that evolve towards the states $\ket{\textsc{ghz}}$ (red solid line) and $\ket{\textsc{w}}$ (blue dashed line) , under the \textsc{dc} and the \textsc{ac} channel, respectively.
  }
  \label{ghzvsw}
\end{figure}


Figure \ref{ghzvsw} depicts the evolution of $\mathbb F_{\max}$ for the initial states 
$\ket{\psi_{\textsc{ghz}}}|_{\phi=\frac{\pi}{4},\varphi=0}$ (red solid line), and $\ket{\psi_{\textsc{w}}}|_{\phi=\arccos\frac{1}{\sqrt 3},\varphi=0}$ (blue dashed line). 
It shows that the maximal average fidelity improves when the resource state involves two qubits from a 3-qubit system that evolves towards the state $\ket{\textsc{ghz}}$, rather than to $\ket{\textsc{w}}$.
%


\section{Fidelity and four-partite entanglement generated via noisy channels}\label{Sfourpartite}

\begin{figure}[ht]
  \begin{center}
    \includegraphics[scale=0.8]{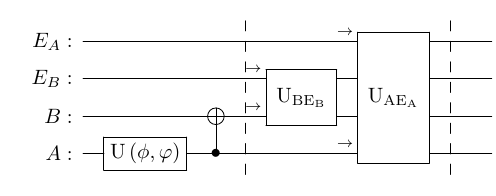}
  \end{center}
  \caption{Qubits $A$, $B$, $E_A$ and $E_B$ are prepared in the state (\ref{initABEE}). The pairs $A,E_A$, and $B,E_B$ interact via unitary operations, thus $A$ and $B$ undergo local channels whose corresponding Kraus operators are $\{Q_\alpha\}$ and $\{K_\beta\}$.}
  \label{esquemaObjT4}
\end{figure}

We now consider that both qubits $A$ and $B$ undergo local channels $\Lambda_{A}$ and $\Lambda_{B}$, as a result of their separate interaction with initially uncorrelated qubits, $E_A$ and $E_B$. 
The initial resource state is thus (see Eqs. (\ref{Irec}) and (\ref{initAB_EAEB})) 
\beq\label{initABEE}
\ket{\psi_0}=\cos\phi \ket{0000}+  e^{i\varphi}\sin\phi \ket{1100},
\eeq
and the Kraus operators of the local channels are generically given by Eq. (\ref{Kraus1}). 
Figure \ref{esquemaObjT4} shows the first two stages of the corresponding teleportation protocol.

Different forms of multipartite entanglement may arise in this 4-qubit system \cite{Amultitangle}. Here we will focus on the multipartite entanglement as measured by the 4-tangle $\tau_4$, defined for a 4-qubit pure state $\ket{\psi}$ as \cite{4tangle} 
\beq\label{tau4}
\tau_{4} =
\vert \bra{\psi} \sigma^{\otimes 4}_{y}\ket{\psi^{*}}\vert^{2}.
\eeq
This quantity becomes maximal ($\tau_4=1$) for the 4-partite \textsc{ghz} state $\frac{1}{\sqrt 2}(\ket{0000}+\ket{1111})$, and vanishes for the 4-partite \textsc{w} state $\frac{1}{\sqrt 4}(\ket{1000}+\ket{0100}+\ket{0010}+\ket{0001})$.  

\subsection{Two local generalized noisy channels}

Assuming that $\Lambda_A$ and $\Lambda_B$ are generalized noisy channels, the corresponding Kraus operators read,  following (\ref{krausACDC}), 
\begin{subequations}\label{kraus2canales}
\begin{equation}\label{Q}
Q_{0}=\left(\begin{array}{cc} 1 & 0 \\ 0 & \sqrt{1-p_{A}} \end{array}\right), \; Q_{1}= \sqrt{p_{A}} \left(\begin{array}{cc} 0 & \cos\zeta_{A} \\ 0 & \sin\zeta_{A} \end{array}\right),
\end{equation}
and
\begin{equation}\label{K}
K_{0}=\left(\begin{array}{cc} 1 & 0 \\ 0 & \sqrt{1-p_{B}} \end{array}\right), \; K_{1}= \sqrt{p_{B}} \left(\begin{array}{cc} 0 & \cos\zeta_{B} \\ 0 & \sin\zeta_{B} \end{array}\right),
\end{equation}
\end{subequations}
with $\zeta_{A(B)} \in[0,\pi/2]$ the parameters that determine the specific channels and $0\leq p_{A(B)}\leq 1$, again with $p_{A(B)}$ appropriate parametrizations of the evolution parameter of the transformation $U_{AE_{A}(BE_{B})}$. 

Resorting to Eq. (\ref{evolstate}) the initial state (\ref{initABEE}) evolves into
\begin{widetext}
\begin{equation}\label{estado4qubits}
 \begin{split}
\ket{\psi}&_{ABE_{A}E_{B}} = \Bigl(\cos\phi\ket{00} + e^{i\varphi}\sin\phi \sqrt{q_{A}q_{B}}\ket{11}\Bigr)\ket{00}
 +e^{i\varphi}\sin\phi  \Big\{  \sqrt{q_{A}p_{B}} \Bigl(\cos\zeta_{B}\ket{10} +\sin\zeta_{B}\ket{11} \Bigr)\ket{01}\\
 +& \sqrt{p_{A}q_{B}} \Bigl(\cos\zeta_{A}\ket{01} +\sin\zeta_{A}\ket{11} \Bigr)\ket{10} 
 +\sqrt{p_{A}p_{B}}  \Bigl(\cos\zeta_{A}\ket{0}+\sin\zeta_{A}\ket{1}\Bigl)\Bigl(\cos\zeta_{B}\ket{0}+ \sin\zeta_{B}\ket{1} \Bigr)\ket{11}\Big\},
 \end{split}   
\end{equation}
\end{widetext}
where we wrote $q_{A(B)}= 1-p_{A(B)}$.
From this state direct calculation gives 
\begin{eqnarray}
\label{4tanglep}
\tau_{4} &=&
 p_{A}p_{B} \big\vert \mathcal E_0 \sin\zeta_{A}\sin\zeta_{B} \nonumber\\ 
&&+\,4e^{i\varphi}
\mathcal P_1
\sqrt{q_{A}q_{B}}\cos\zeta_{A}\cos\zeta_{B} \big\vert^{2}.
\end{eqnarray}

Unlike the previous case in
which only $B$ passes through the generalized channel, where $C_{AB}$ was independent of the channel parameter $\zeta$ (see Eq. (\ref{concuACDC})), in the present scenario the entanglement of the resource state depends on both $\zeta_A$ and $\zeta_B$. 
However, the general and explicit dependence of $C_{AB}$ on $\zeta_A,\zeta_B$ is far from trivial and will be omitted here. 
It suffices to recall that $C_{AB}$ can be obtained from Eq. (\ref{concumixed}) using the resource state $\rho_{AB}=\textrm{Tr}_{E_{A}E_{B}}\ket{\psi}\!\bra{\psi}$, with  $\ket{\psi}$ given by (\ref{estado4qubits}).

From Eq. (\ref{fmaxKBlocal}) the maximal average fidelity for two \textsc{gc} channels is $F_{\max}=\max_{\,i}\,\{F_{\Phi_{i}}\}$ 
with 
\begin{equation}\begin{split}
F_{\Phi_i}&=\frac{1}{3} + \frac{2}{3}  \Big\{ \sum_{\alpha\beta} \big\vert \langle\Phi_i\big\vert Q_{\alpha}\otimes K_{\beta} \big\vert\phi_{0}\rangle  \big\vert^{2} \Big\}.
\end{split}
\end{equation} 
Substitution of the Kraus operators (\ref{kraus2canales}) gives 
\begin{subequations}
\begin{eqnarray}
\label{fidchorizo}
F_{\Phi^{\pm}}&=&\frac{2}{3} +\frac{1}{3}\Big(\pm \mathcal{E}_{0}\sqrt{q_{A}q_B}\cos{\varphi} - \mathcal{P}_{1}\Delta_{\mp}\Big),\\
F_{\Psi^{\pm}}&=&\frac{2}{3}-\frac{1}{3}\Big(1-\mathcal{P}_{1}\Delta_{\pm}\Big),
\end{eqnarray}
\end{subequations}
where $\Delta_{\pm}\in[0,1]$ and is given by
\begin{eqnarray}
\Delta_{\pm}&=&  q_Bp_{A}\cos^{2}\zeta_{A}+q_Ap_{B}\cos^{2}\zeta_{B}\nonumber\\
&&+\,p_{A}p_{B}\sin^{2}(\zeta_{A} \pm\zeta_{B}).
\end{eqnarray}
Consequently  $F_{\Psi^{\pm}} \leq 2/3$; further, since $\zeta_{A},\zeta_{B} \in [0,\pi/2]$ it holds that  $\sin^{2}(\zeta_{A}+\zeta_{B}) \geq \sin^{2}(\zeta_{A}-\zeta_{B})$, whence $\Delta_{+}\geq \Delta_{-}$.
This in turn leads us to
\beq
\cos \varphi\geq 0 \Rightarrow F_{\Phi^{+}}\geq F_{\Phi^{-}},
\eeq
so in the present case $\mathbb F_{\max}$, given by Eq. (\ref{fmaxnos}), writes as
\beq
\label{maf4qubit0}
\mathbb{F}^{\textsc{gc/gc}}_{\max \pm}=\frac{2}{3} +\frac{1}{3}
\max\Big\{0,
\Big[\mathcal{E}_{0}\sqrt{q_{A}q_B}\,\vert\!\cos{\varphi}\vert -\mathcal P_1 \Delta_{\mp}\Big]
\Big\}
\eeq
%
where the upper/lower sign must be chosen accordingly with 
\beq\label{descoseno}
\cos\varphi  \gtrless -\frac{1}{4}\frac{\sqrt{\mathcal P_1}}{\sqrt{1-\mathcal P_1}}\frac{1}{\sqrt{q_Aq_B}}(\Delta_+-\Delta_-).
\eeq
As follows from (\ref{maf4qubit0}), the optimal phase $\varphi$ is the same as in the previous cases, namely $\varphi=0,\pi$. Also, a loss in the fidelity emerges, encoded in the negative term proportional to $\mathcal P_1$.


 \subsection{Fidelity and multipartite entanglement for 4-qubits under parallel generalized noisy channels}

In our forthcoming analysis, we will simplify the equations  taking $p_{A}=p_{B}=p$ 
\footnote{This means that the evolution parameters of $U_{AE_A}$ and $U_{BE_B}$ are the same (for example they are both the time $t$), and the parametrizations are identical, so $p(t)$ is the same for both pairs of Kraus operators (\ref{kraus2canales}).}.
Equation (\ref{4tanglep}) thus reads 
\begin{eqnarray}
\label{4tanglepB}
\tau_{4} &=&p^2\Big[\mathcal E_0\sin\zeta_A\sin\zeta_B+4\mathcal P_1(1-p)\cos\zeta_A\cos\zeta_B\Big]^2\nonumber\\
&&-2\mathcal E_0\mathcal P_1 p^2(1-p)\sin 2\zeta_A\sin 2\zeta_B (1-\cos\varphi).
\end{eqnarray}
Recalling that $\zeta_{A,B}\in[0,\pi/2]$, the second line in Eq. (\ref{4tanglepB}) is a non-positive term that reduces the amount of 4-partite entanglement. 
 Notice, however, that for $\varphi=0$ such term vanishes, and consequently the relative phase that maximizes $\tau_4$ maximizes also $\mathbb{F}^{\textsc{gc/gc}}_{\max \pm}$. 
 By choosing this optimal phase, Eq. (\ref{descoseno}) holds with the upper inequality sign, and accordingly Eq. (\ref{maf4qubit0}) reduces to 
 \begin{eqnarray}
\label{fidchorizo1}
\mathbb{F}^{\textsc{gc/gc}}_{\max}&=&\frac{2}{3} +\frac{1}{3}\max\Big\{0,\Big[ \mathcal{E}_{0}(1-p)-\mathcal P_1 p[p\sin^2(\zeta_A-\zeta_B)\nonumber\\
&& 
+(1-p) (\cos^2\zeta_A+\cos^2\zeta_B)]
\Big]\Big\},
\end{eqnarray}
 %
whereas $\tau_4$ becomes 
\beq
\label{4tanglepBb}
\tau_{4} =\Big[\mathcal E_0p\sin\zeta_A\sin\zeta_B+4\mathcal P_1p(1-p)\cos\zeta_A\cos\zeta_B\Big]^2.
\eeq

If $\mathcal E_0p\geq4\mathcal P_1p(1-p)$ then
\beq
\mathcal E_0p \cos\zeta_A\cos\zeta_B\geq 4\mathcal P_1p(1-p)\cos\zeta_A\cos\zeta_B,
\eeq
whence, by adding $\mathcal E_0p \sin\zeta_A\sin\zeta_B$ on both sides of the inequality we get
\beq
\mathcal E_0p \cos(\zeta_A-\zeta_B)\geq\sqrt{\tau_4}.
\eeq
An analogous reasoning applies for $\mathcal E_0p\leq4\mathcal P_1p(1-p)$, so we finally get
\beq
\tau_4\leq T^2\cos^2(\zeta_A-\zeta_B),
\eeq
where $T=\max\{\mathcal E_0p, 4 \mathcal{P}_1 p
 (1-p)\}$. 
 This means that the maximum value of $\tau_4$ over all possible channels is located along the line $\zeta_A=\zeta_B=\zeta$, and explicitly reads
\beq \label{maxtau4}
\tau^{\max}_{4}=\begin{cases}
			\mathcal E^2_0p^2=\tau_4|_{\zeta=\pi/2} \!& \textrm{if}\;\, \mathcal E_0\geq4 \mathcal{P}_1 
 (1-p),\\
            16 \mathcal{P}^2_1 
 p^2(1-p)^2= \tau_4|_{\zeta=0} \!& \textrm{if}\;\, \mathcal E_0\leq4 \mathcal{P}_1 
 (1-p).
		 \end{cases}
\eeq
The maximum 4-partite entanglement is thus reached whenever both channels are either \textsc{dc} or \textsc{ac}. 
In the former case also  $\mathbb{F}^{\textsc{gc/gc}}_{\max}$ reaches its maximum value, over all \textsc{gc/gc} channels. 
This can be seen by noticing that the loss in the fidelity, encoded in the  (non-positive) term proportional to $\mathcal P_1p$ in Eq. (\ref{fidchorizo1}), vanishes irrespective of the initial state and the evolution parameter only for $\zeta_{A}=\zeta_B=\pi/2$, resulting in
\begin{eqnarray}\label{Fdcdc}
\mathbb{F}^{\textsc{gc/gc}}_{\max}\leq \mathbb{F}^{\textsc{dc/dc}}_{\max}&=&
\frac{2}{3} +\frac{1}{3}\mathcal{E}_{0}(1-p)\\
&=& \frac{2}{3} +\frac{1}{3}C^{\textsc{dc/dc}}_{AB}(p),\nonumber
\end{eqnarray}
where $C^{\textsc{dc/dc}}_{AB}=\mathcal{E}_{0}(1-p)$ stands for the entanglement of the resource state when two parallel \textsc{dc} channels are implemented.
Consequently, whenever the condition in the first line of (\ref{maxtau4}) holds, i.e., whenever 
 \beq\label{cond_ft}
 1-\frac{1}{2}\sqrt{\frac{1-\mathcal P_1}{\mathcal P_1}}\leq p,
\eeq
the channel that maximizes $\tau_4$ 
 maximizes also $\mathbb{F}^{\textsc{gc/gc}}_{\max}$.  
Notice that (\ref{cond_ft}) is satisfied for all $p$ provided $\mathcal P_1\leq 1/5$. That is, for sufficiently low populations $\mathcal P_1$, the  maximal value of $\tau_4$ and of $\mathbb{F}^{\textsc{gc/gc}}_{\max}$ are jointly reached. 

If the condition (\ref{cond_ft}) is not met, $\tau^{\max}_4$ is given by the second line in Eq. (\ref{maxtau4}) and corresponds to $\zeta_A=\zeta_B=0$, i.e., to the combination \textsc{ac/ac}. In this case (\ref{fidchorizo1}) reduces to
\begin{eqnarray}
\label{fid_acac}
\mathbb{F}^{\textsc{ac/ac}}_{\max}&=&\frac{2}{3} +\frac{1}{3}C^{\textsc{ac/ac}}_{AB}(p)\\
&=&\max\Big\{\frac{2}{3},\mathbb{F}^{\textsc{dc/dc}}_{\max}-\frac{2}{3}\mathcal P_1 p(1-p)\Big\},\nonumber
\end{eqnarray}
where in the first line we used that under parallel \textsc{ac}, $C^{\textsc{ac/ac}}_{AB}=\max\{0, (1-p) (\mathcal E_0-2\mathcal P_1 p)\}$.

Now, from Eq. (\ref{fidchorizo1}) we obtain the following expression for $\mathbb{F}^{\textsc{gc/gc}}_{\max}$ along the line $\zeta_A=\zeta_B=\zeta$:
 \begin{eqnarray}
\label{fidchorizo1_id}
\mathbb{F}^{\textsc{gc/gc}}_{\max}(\zeta)&=&\frac{2}{3} +\frac{1}{3}\max \Big\{0,\Big[ \mathcal{E}_{0}(1-p)\nonumber\\
&&-2\mathcal P_1 p
(1-p)\cos^2\zeta
\Big]\Big\},
\end{eqnarray}
%
whose minimum above the threshold value $2/3$ corresponds to the channel \textsc{ac/ac}. 
This observation, together with the fact that $\tau^{\max}_4$ is attained at $\zeta_A=\zeta_B=\zeta=0,\pi/2$, indicates that 
the channels that maximize the 4-partite entanglement are not always those that maximize $\mathbb{F}^{\textsc{gc/gc}}_{\max}$.
Rather,  $\mathbb{F}^{\textsc{gc/gc}}_{\max}$ is either maximal or minimal (within the family of twin channels with $\zeta_A=\zeta_B=\zeta$) on  points where the 4-tangle is maximal.

As for the vanishing value of $\tau_4$, we see from Eq. (\ref{4tanglepBb}) that it corresponds to the combination of a \textsc{dc} on either one of the qubits, and an \textsc{ac} on the other one. This null 4-tangle is accompanied by the following maximal average fidelity 
%
\begin{eqnarray}
\label{Fcruzada}
\mathbb{F}^{\textsc{ac/dc}}_{\max}=\mathbb{F}^{\textsc{dc/ac}}_{\max}&=&\frac{2}{3} +\frac{1}{3}\max\Big\{0, \mathcal{E}_{0}(1-p)-\mathcal P_1 p
\Big\}\nonumber\\
&=&\max\Big\{\frac{2}{3},\mathbb{F}^{\textsc{dc/dc}}_{\max}-\frac{1}{3}\mathcal P_1 p\Big\},
\end{eqnarray}
an expression that is analogous to Eq. (\ref{FmaxAC}) for the case with vanishing 3-tangle (\textsc{ac}).\\
\begin{figure*}
    \includegraphics[scale=0.65]{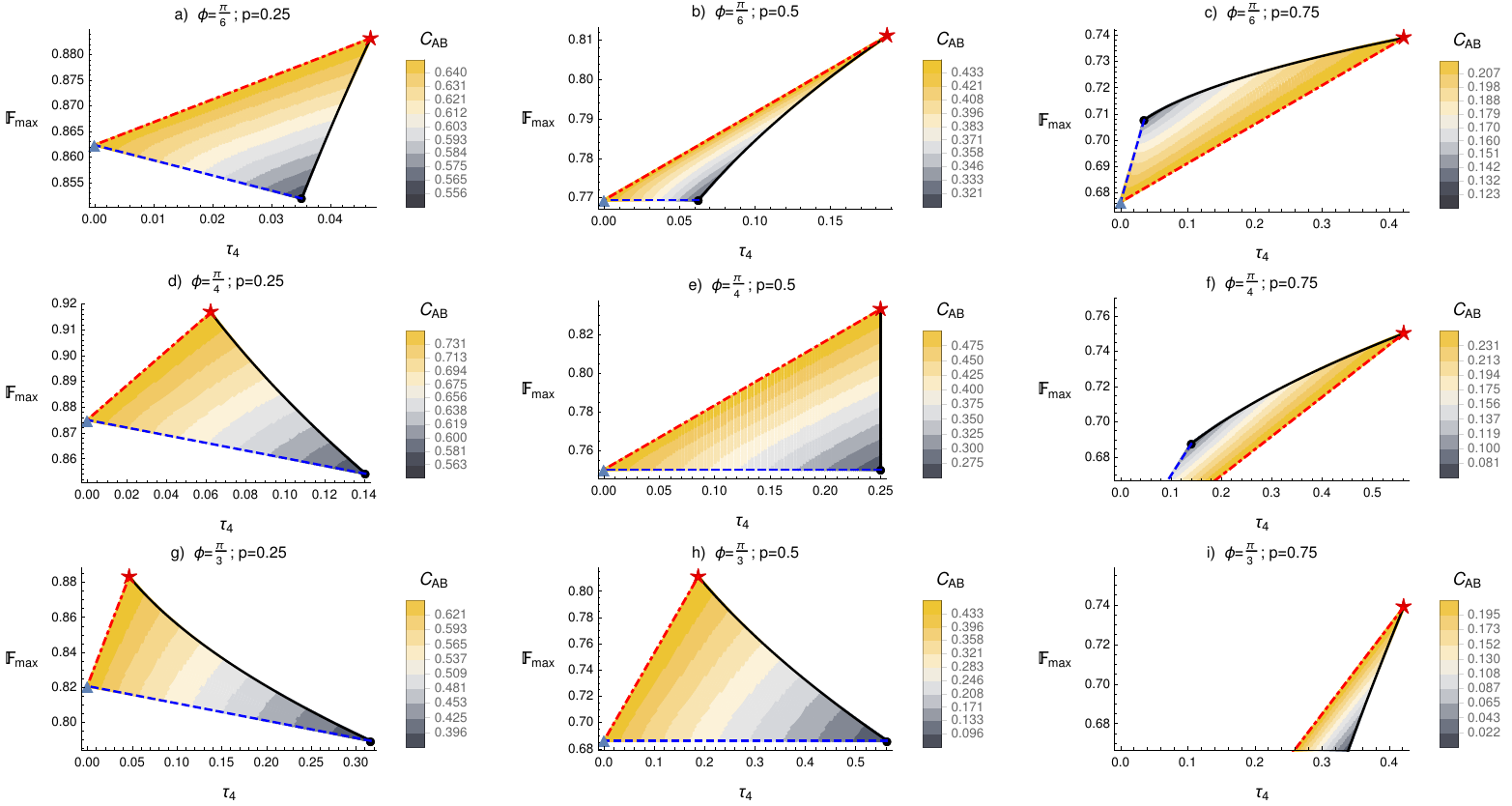}
  \caption{$\mathbb{F}^{\textsc{gc/gc}}_{\max}$ versus $\tau_4$ for different values of $\phi$ (rows, from top to bottom: $\phi=\pi/6,\pi/4,\pi/3$) and $p$ (columns, from left to right: $p=0.25, 0.5, 0.75$). 
  The regions are generated as all the channels are considered (by varying $\zeta_A$, $\zeta_B$), and the color scale indicates the amount of entanglement of the corresponding resource state, $C_{AB}$.  
  The confining curves comprise different families of channels: twin channels with $\zeta_A=\zeta_B=\zeta$ along the solid black curve; \textsc{ac/gc} (or \textsc{gc/ac}) channels along the dashed blue curve; \textsc{dc/gc} (or \textsc{gc/dc}) channels along the dashed-dotted red curve.  
  Black circles correspond therefore to $\zeta_A=\zeta_B=0$ (parallel \textsc{ac}), red stars to $\zeta_A=\zeta_B=\pi/2$ (parallel \textsc{dc}), and blue triangles to $\zeta_{A(B)}=0,\zeta_{B(A)}=\pi/2$ (hybrid \textsc{ac/dc}, \textsc{dc/ac}).  }
  \label{corrmap}
\end{figure*}

In order to make a complete analysis of the relation between $\tau_4$ and $\mathbb{F}^{\textsc{gc/gc}}_{\max}$, a key element should be considered, namely the entanglement of the resource state.
The evolution of $C_{AB}$ and $\mathbb{F}^{\textsc{gc/gc}}_{\max}$ for a given initial state and fixed channel parameters is qualitatively similar to that shown in Fig. \ref{f3qubitevol}, i.e., the quantum channel  induces a loss in $C_{AB}$, which is verified by noticing that Eq. (\ref{fidchorizo1})  is a decreasing function of $p$.
Yet, as occurred in the tripartite case, there exists a positive correlation between $\mathbb{F}^{\textsc{gc/gc}}_{\max}$ and the multipartite entanglement (here measured by $\tau_4$), that holds at each stage of the evolution, and for a fixed amount of $C_{AB}$.
In order to verify this, it becomes crucial to recall first 
an important difference between the 3- and the 4-partite scenarios: 
as mentioned below Eq. (\ref{4tanglep}), in the former case $C_{AB}$ is independent of $\zeta$ (hence is the same for all channels), whereas in the 4-qubit example $C_{AB}$ typically varies with both channel parameters, $\zeta_A$ and $\zeta_B$.
This observation, together with the essential role played by the entanglement of the resource state in the teleportation success, calls for an analysis that incorporates the value of $C_{AB}$ when examining the dynamic interplay between $\tau_4$ and $\mathbb{F}^{\textsc{gc/gc}}_{\max}$.
To this end, we compare the effect of various channels (scanned by varying $\zeta_A$ and $\zeta_B$) on the quantities of interest, namely $\tau_4, \mathbb{F}^{\textsc{gc/gc}}_{\max}$ and $C_{AB}$, at the same stage of the evolution for different initial states. 
We do so by constructing triads $(\tau_4, \mathbb{F}^{\textsc{gc/gc}}_{\max}, C_{AB})$, which for a given $p$ and fixed initial parameters $\varphi$ and $\phi$, depend only on $\zeta_A,\zeta_B$. 
We then consider 16,384 different channels (pairs $\{\zeta_A,\zeta_B\}$), 
and display the resulting triads as points in the plane $(\tau_4,\mathbb{F}^{\textsc{gc/gc}}_{\max})$ colored according to the corresponding range of values of $C_{AB}$.
Figure \ref{corrmap} shows the ensuing triads for $\varphi=0$ and different values of $\phi$ (rows) and $p$ (columns).
In all cases, $\tau_4$ and $\mathbb{F}^{\textsc{gc/gc}}_{\max}$ were directly computed from Eqs. (\ref{4tanglepBb}) and (\ref{fidchorizo1}), respectively, whereas $C_{AB}$ was numerically calculated as explained below Eq. (\ref{4tanglep}).
%

All points in each panel of Fig. \ref{corrmap} lie within a region delimited by three curves:
\begin{itemize}
\item The black solid line, encompassing the family of twin channels with $\zeta_A=\zeta_B=\zeta$.

\item The blue dashed line, including those cases where one of the qubits (either $A$ or $B$) undergoes an \textsc{ac}, meaning $\zeta_{A/B}=0$, while the other one undergoes an arbitrary \textsc{gc}.

\item  The red dotted-dashed line, comprising the channels where one of the qubits (either $A$ or $B$) undergoes a \textsc{dc}, so $\zeta_{A/B}=\pi/2$, while the other one is subject to an arbitrary \textsc{gc}.
\end{itemize}
Accordingly, the vertices of the regions are identified as:
\begin{itemize}
\item  Blue  triangles represent the \textsc{ac/dc} (\textsc{dc/ac}) channels, with $\tau_4=0$ and maximal average fidelity given by (\ref{Fcruzada}).

\item Red stars correspond to \textsc{dc/dc}, where  $\mathbb{F}^{\textsc{gc/gc}}_{\max}$ reaches its maximum value, in line with Eq. (\ref{Fdcdc}). 

\item Black circles represent the channel \textsc{ac/ac},  and correspond to points with the lowest $\mathbb{F}^{\textsc{gc/gc}}_{\max}$  along the line of twin channels, in agreement with the statement below Eq. (\ref{fidchorizo1_id}).
\end{itemize}

In all the graphs in Fig. \ref{corrmap}, the colored bands, containing points whose $C_{AB}$ lies within a specific range of values, reveal a positive correlation between the 4-tangle and the maximal average fidelity. 
Such correlation persists as the width of the range is reduced, as exemplified in Table \ref{tab:totopo} showing values of  
$C_{AB}$, $\tau_4$ and  $\mathbb{F}^{\textsc{gc/gc}}_{\max}$ taken from the central panel in Fig. \ref{corrmap}.
The values of  $\tau_4$ and  $\mathbb{F}^{\textsc{gc/gc}}_{\max}$ are displayed in increasing order, and both increase simultaneously as the value of $C_{AB}$ (numerically obtained) remains constant (up to 4 digits).
The results indicate that despite the specificities present in each panel of Fig. \ref{corrmap}, a feature common to all of them is that \emph{for a fixed entanglement of the resource state} $C_{AB}$, $\mathbb{F}^{\textsc{gc/gc}}_{\max}$ increases as $\tau_4$ increases.
\begin{table}[h]
    \centering
    \begin{tabular}{|c|c|c|c|c|}
    \hline
      $\zeta_{A}/\pi$  & $\zeta_{B}/\pi$  &  $C_{AB}$& $\tau_{4}$  & $\mathbb{F}^{\textsc{gc/gc}}_{\max}$  
      \\
      \hline
      \hline
       181/500 & 37/500 & 0.422003 & 0.0954376 & 0.760765 
       \\ \hline 
       91/250 & 2/25 & 0.422002 & 0.0984991 & 0.761839
       \\ \hline 
       93/250 &  29/250 & 0.422003 & 0.120289 & 0.768959
       \\ \hline
       187/500 & 143/1000 & 0.422001 & 0.139887 & 0.774978
       \\ \hline 
       187/500 & 19/125 & 0.422008 & 0.146878 & 0.777085
       \\ \hline
       369/1000 & 99/500 & 0.422009 & 0.18453 & 0.788234
       \\ \hline 
       42/125 & 281/1000 & 0.422005 & 0.24261 & 0.805185
       \\ \hline
       167/500 & 71/250 & 0.422008 & 0.243882 & 0.805556
       \\ \hline
    \end{tabular}
    \caption{Values of 
$C_{AB}$, $\tau_4$ and $\mathbb F^{\textsc{gc/gc}}_{\max}$  extracted from the panel e) of Fig. \ref{corrmap} 
 (corresponding to $\phi=\pi/4$, $\varphi=0$, and $p=0.5$), for different values of the channels parameters $\zeta_A$ and $\zeta_B$.}
    \label{tab:totopo}
\end{table}
In other words, for a given initial state ($\phi$ fixed), and at each stage of the evolution ($p$ fixed), from among all the channels \textsc{gc/gc} that correspond to the same entanglement of the resource state, those that generate higher amounts of 4-partite entanglement lead to higher fidelities. 
This conclusion is analogous to that reached in the 3-partite case, indicating that the multipartite entanglement may act as a resource that protects the teleportation fidelity against the noisy channel (provided the same amount of $C_{AB}$ is at disposal).

For the values of $\phi$ and $p$ considered in panels a), b), c),  f), and i) the condition (\ref{cond_ft}) is satisfied, and accordingly the maximum of $\mathbb{F}^{\textsc{gc/gc}}_{\max}$ is reached along with the maximum of the 4-tangle, given in this case by $\tau^{\max}_4=\mathcal E^2_0 p^2$.
In contrast, the cases depicted in panels d), g) and h) do not comply with the inequality (\ref{cond_ft}), thus correspond to $\tau^{\max}_4=16\mathcal P^2_1 p^2(1-p)^2$ and $\mathbb{F}^{\textsc{gc/gc}}_{\max}=\mathbb{F}^{\textsc{ac/ac}}_{\max}$, that is, to the minimal fidelity along the black solid curve. 
Panel e) corresponds to the case in which $\mathcal E_0=4\mathcal P_1 (1-p)$, so as follows from Eq. (\ref{maxtau4})  the maximum value of $\tau_4$ is attained simultaneously at the red star (\textsc{dc/dc}) and the black circle (\textsc{ac/ac}).

In all panels of Fig. \ref{corrmap} it is observed that as the black solid curve is traversed from the black circle to the red star, $C_{AB}$ increases along with $\mathbb{F}^{\textsc{gc/gc}}_{\max}$, so the highest fidelity is reached along with the highest $C_{AB}$.
This shows that, within the family of twin channels, the resource state's entanglement helps to improve the maximal average fidelity. 
However, this does not hold for other channels. 
For example, as the dashed blue line is traversed from the black circle to the blue triangle ---that is, along the family of channels \textsc{ac/gc}---, there are cases in which an increase in $C_{AB}$ is accompanied by a decrease in $\mathbb{F}^{\textsc{ac/gc}}_{\max}$, as the one depicted in panel c). 
In such case it is only the 4-tangle what is seemingly enhancing the maximal average fidelity. 
Further, for the family of channels \textsc{dc/gc}, as the red dotted-dashed line is traversed, $C_{AB}$ is kept constant (for fixed $p$) and maximal ---in fact, in this case we get $C^{\textsc{dc/gc}}_{AB}=\mathcal{E}_0(1-p)$---, and the improvement of $\mathbb{F}^{\textsc{dc/gc}}_{\max}$ is therefore ascribable to the increase of $\tau_4$.


\section{Closing remarks}\label{Close}

We investigated the role of multipartite entanglement in the dynamics of the maximal average fidelity (above the classical threshold value) when the teleportation protocol includes the action of a local quantum channel, $\Lambda_{AB}=\Lambda_{A}\otimes \Lambda_{B}$, acting on the qubits $A$ and $B$ that conform the resource state. 
To facilitate our goal we expressed the maximal average fidelity in terms of the Kraus operators corresponding to a general 2-qubit channel (Eq. (\ref{fmaxKB})), and introduced the Kraus operators of a generalized noisy channel, Eq. (\ref{krausACDC}), which encompasses the paradigmatic amplitude damping and dephasing channels, and connects them via a continuous parameter that also determines the amount of multipartite entanglement created along the evolution. 
%

We first considered the case $\Lambda_{A}=\mathsf I_2$, and $\Lambda_{B}$  representing the generalized noisy channel, rooted at the interaction of $B$ with an additional qubit $E_B$.
 3-partite entanglement thus typically emerges among the qubits $A, B$, and $E_B$, here quantified by the 3-tangle $\tau_{ABE_B}$. 
In the second scenario, both $\Lambda_{A}$ and $\Lambda_{B}$ correspond to generalized channels. 
Interpreting them as the effective result of a local interaction of $A$ and $B$ with additional qubits $E_A$ and $E_B$, we focused on the ensuing 4-partite entanglement among the parties $A, B, E_A$, and $E_B$, as measured by $\tau_4$. 

In both cases, we found that the relative phase $\varphi$ that optimizes both $\mathbb F_{\max}$ and $\tau_4$ is $\varphi=0$. 
More importantly,  the analytical and numerical analysis (considering identically parametrized channels in the 4-party case),
revealed that 
\emph{for a fixed non-zero amount of the resource state's entanglement (i.e., for fixed $C_{AB}\neq 0$), and at each stage of the evolution, the teleportation success improves under channels that induce higher amounts of multipartite entanglement.}
Here it is important to stress that both the 3-tangle and the 4-tangle quantify a specific type of multipartite entanglement ---namely the 3-way and the 4-way entanglement, respectively--- characteristic of \textsc{ghz}-type states, and thereby absent in \textsc{w}-type states \cite{DurPRA2000,4tangle}. 
Consequently, our findings indicate that it is specifically the amount of \textsc{ghz}-type entanglement which favors better teleportation fidelities.
It should be stressed that this conclusion does not go against the (expected) fact that, 
given a specific noisy channel (a fixed $\zeta$), as the evolution takes place  the teleportation fidelity decays while the 3- and 4-tangle may increase;
rather, the conclusion compares the effect of \emph{different} generalized noisy channels on $\mathbb F_{\max}$ throughout the evolution, and establishes that under channels that produce more multipartite (\textsc{ghz}-type) entanglement, the detrimental effects on $\mathbb{F}_{\max}$ are lessened. 
Further,
when $C_{AB}$ vanishes, as occurs at $p=1$ ---when the channel has suppressed all the entanglement of the resource state---, the maximal average fidelity
drops below the classical threshold value, despite the multipartite entanglement may acquire relatively large values. 
This highlights $C_{AB}$ as a necessary element that triggers the power of the $n$-way entanglement to enhance the teleportation success. 


In the 3-party case, $C_{AB}$ is the same for all the channels considered, and the relation between $\mathbb F^{\textsc{gc}}_{\max}$ and $\tau_{ABE_B}$ was clearly revealed, along with the
identification of the \textsc{dc} as the channel that produces the higher values of $\mathbb F^{\textsc{gc}}_{\max}$. 
In the 4-qubit case, in contrast, $C_{AB}$ depends on the specific channel, and the relation between $\mathbb F^{\textsc{gc/gc}}_{\max}$ and $\tau_4$ \emph{only} (i.e., without considering the value of $C_{AB}$) is more subtle than in the 3-party counterpart. 
In particular, we found that an increment in $\tau_4$ induces an increment in $\mathbb F^{\textsc{gc/gc}}_{\max}$, irrespective of the initial state and stage of the evolution, only when the composite channel is \textsc{dc/gc}, that is, when either one of the qubits is subject to a dephasing channel. Notably, in this case, $C_{AB}$ does not depend on the specific \textsc{gc}, and the enhancement of the maximal average fidelity is due solely to the increase in the 4-partite entanglement.

For the families of twin channels \textsc{gc/gc} with $\zeta_A=\zeta_B=\zeta$, and \textsc{ac/gc}, a higher value of $\tau_4$ is not always accompanied by a higher value of $\mathbb F^{\textsc{gc/gc}}_{\max}$, ultimately because $C_{AB}$ changes with the channels' parameters. 
Instead, for a fixed initial state and at a given stage of the evolution, the composite channel for which 
the 4-tangle is maximal is either \textsc{dc/dc} ---in which case $\mathbb F^{\textsc{gc/gc}}_{\max}$ attains its global \emph{maximal} value---, or  \textsc{ac/ac}, corresponding to a $\mathbb F^{\textsc{gc/gc}}_{\max}$ that is \emph{minimal}   within the family of twin channels ($\zeta_A=\zeta_B=\zeta$).  

Interestingly, guaranteeing that $\mathbb F_{\max}$ exceeds the classically attainable value 
depends not only on the initial entanglement at disposal, but also on the initial population $\mathcal P_1$ of the state $\ket{11}_{AB}$, which plays against the improvement of the teleportation success. 
Further, for sufficiently low values of $\mathcal P_1$ ($\mathcal P_1\leq 0.2$) the maximal value of $\tau_4$ is attained together with the maximal value of $\mathbb F^{\textsc{gc/gc}}_{\max}$.

Our analysis led us to conclude that the previously reported improvement of the teleportation fidelity under some types of noisy channels \cite{badzia2000,BandyoPRA2002,Fortes2015,Ahadpour2020,FonsecaPRA2019,Laura2014} may be rooted at the emergence of multipartite entanglement induced by the interaction of the resource qubits with their surroundings. This offers valuable insights into the power of multipartite correlations, 
as well as into the characterization of the processes that better protect the teleportation fidelity in the more realistic scenario in which $A+B$ is an open system. 
In particular, processes that generate \textsc{ghz}-type states in the 3- or 4-qubit system have the potential to assist the protocol by reducing the detrimental effects of noise, as a result of the induced generation of 3- and 4-way entanglement. This highlights the \textsc{ghz}-type entanglement as a useful auxiliary resource in noisy quantum teleportation.

\begin{acknowledgments}
The authors acknowledge financial support from DGAPA, UNAM through project PAPIIT IN112723. VHTB acknowledges CONAHCYT scholarship with CVU: 863195.
\end{acknowledgments}


\end{document}